\documentclass[letterpaper]{article} 
\usepackage{newunicodechar}
\newunicodechar{₇}{$_7$}
\usepackage{aaai25}  
\usepackage{times}  
\usepackage{helvet}  
\usepackage{courier}  
\usepackage[hyphens]{url}  
\usepackage{graphicx} 
\usepackage{natbib}  
\usepackage{caption} 
\usepackage{algorithm}
\usepackage{algorithmic}

\usepackage{newfloat}
\usepackage{listings}
\DeclareCaptionStyle{ruled}{labelfont=normalfont,labelsep=colon,strut=off} 
\floatstyle{ruled}
\newfloat{listing}{tb}{lst}{}
\floatname{listing}{Listing}
\usepackage{xcolor}
\newcommand{\answerYes}[1]{\textcolor{blue}{#1}} 
 
\newcommand{\answerNA}[1]{\textcolor{gray}{#1}} 
\usepackage{booktabs}
\usepackage{array}
\usepackage{soul}
\usepackage{cellspace}
\newcommand{\hlc}[2][yellow]{{%
    \colorlet{foo}{#1}%
    \sethlcolor{foo}\hl{#2}}%
}

\usepackage{multirow}
\newcommand\blfootnote[1]{%
  \begingroup
  \renewcommand\thefootnote{}\footnote{#1}%
  \addtocounter{footnote}{-1}%
  \endgroup
}

\title{Unifying the Extremes: Developing a Unified Model for Detecting and Predicting Extremist Traits and Radicalization}
\author {
    Allison Lahnala\equalcontrib\textsuperscript{\rm 1},
    Vasudha Varadarajan\equalcontrib\textsuperscript{\rm 2},
    Lucie Flek\textsuperscript{\rm 3},
    H. Andrew Schwartz\textsuperscript{\rm 2},
    Ryan L. Boyd\textsuperscript{\rm 4}
}
\affiliations {
    \textsuperscript{\rm 1}Department of Computing \& Software, McMaster University\\ 
    \textsuperscript{\rm 2}Department of Computer Science, Stony Brook University\\
    \textsuperscript{\rm 3}Bonn-Aachen International Center for Information Technology, University of Bonn\\
    \textsuperscript{\rm 4}Department of Psychology, University of Texas at Dallas
}

\usepackage{bibentry}

\begin{document}

\maketitle
\blfootnote{Corresponding Author: Ryan L. Boyd, boyd@utdallas.edu}
\begin{abstract}
 The proliferation of ideological movements into extremist factions via social media has become a global concern. While radicalization has been studied extensively within the context of specific ideologies, our ability to accurately characterize extremism in more generalizable terms remains underdeveloped. In this paper, we propose a novel method for extracting and analyzing extremist discourse across a range of online ideological community forums. By focusing on verbal behavioral signatures of extremist traits, we develop a framework for quantifying extremism at both user and community levels. Our research identifies 11 distinct factors, which we term ``The Extremist Eleven,'' as a generalized psychosocial model of extremism. Applying our method to various online communities, we demonstrate an ability to characterize ideologically diverse communities across the 11 extremist traits. We demonstrate the power of this method by analyzing user histories from members of the incel community. We find that our framework accurately predicts which users join the incel community up to 10 months before their actual entry with an AUC of $>0.6$, steadily increasing to AUC $\sim0.9$ three to four months before the event. Further, we find that upon entry into an ideological forum, the users tend to maintain their level of extremist traits within the community, while still remaining distinguishable from the general online discourse. Our findings contribute to the study of extremism by introducing a more holistic, cross-ideological approach that transcends traditional, trait-specific models.
 \end{abstract}

%
\begin{links}
    \link{Code}{https://github.com/humanlab/extremism}
\end{links}

\section{Introduction}

\textcolor{red}{Sensitive content warning: This paper may contain offensive language.} The proliferation of extremist ideologies in online spaces has become a pressing issue in recent years, with widespread implications for societal stability and individual well-being. Extremism, broadly defined by its advocacy for rigid and uncompromising ideologies, has influenced political, social, and cultural discourse in profound ways. These movements often promote exclusionary or radical views that challenge the prevailing norms of society, sometimes advocating for the use of violence or extreme measures to achieve their objectives~\cite[e.g.,][]{borum_radicalization_2011}. As extremist content becomes increasingly accessible through digital platforms, understanding its underlying drivers is crucial not only for addressing its spread but also for mitigating the potential harm it may cause~\cite{stephens_preventing_2021}.
\begin{figure}[t!]
    \centering
    \includegraphics[width=0.9\linewidth]{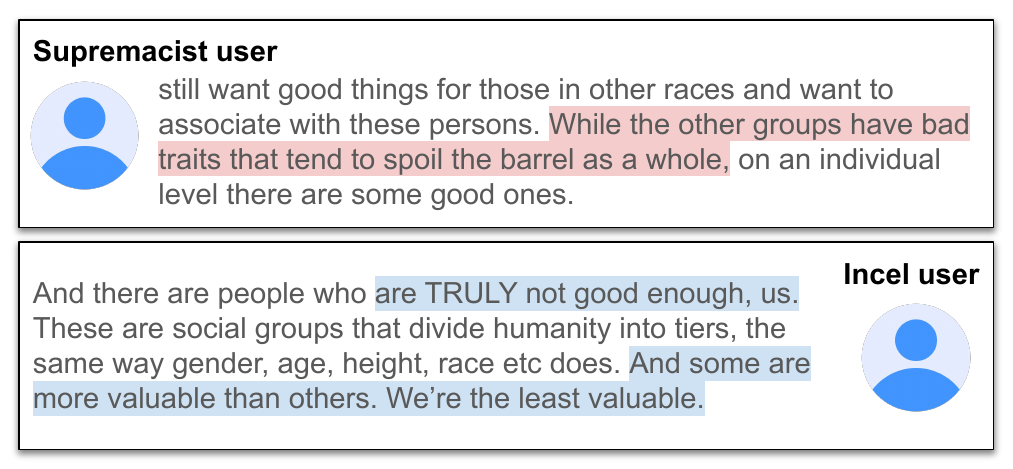}
    \caption{Two very different extremist narratives can still have underlying commonalities. These messages from people holding distinct extremist beliefs reveal a common underlying belief in social hierarchy, despite one exhibiting supremacy and the other exhibiting inferiority. Our work explores the larger facets of extremism by integrating and studying multiple ideologies together to understand the broad differences and similarities that can characterize each kind of extremism.  }
    \label{fig:spirit}
\end{figure}
The rise of online extremist communities has also created spaces for recruitment, radicalization, and the formation of social processes that reinforce harmful ideologies~\cite{tornberg_echo_2024}. These digital environments enable like-minded individuals to connect, reinforcing shared grievances and amplifying extremist rhetoric. The accessibility and anonymity of the internet provide a fertile ground for the proliferation of these movements, making it vital to understand how these communities function and why they are so effective in recruiting new members. This understanding is key to informing strategies aimed at preventing radicalization and countering the influence of these groups.

One of the challenges in studying extremism is the diversity of groups and ideologies that fall under this umbrella~\cite[e.g.,][]{doering_reconceptualizing_2023}. Whether rooted in ethno-nationalism, radical religious ideologies, misogyny, or anti-capitalist sentiment, these groups often differ in their stated goals and motivations (see Figure~\ref{fig:spirit}). However, they frequently share underlying psychological and social patterns~\cite{gartenstein-ross_composite_2023}, such as perceived marginalization \cite{hales_marginalized_2018}, grievances with societal structures \cite{kearns_political_2020}, and promotion of drastic societal change in many forms, ranging from anarchic dismantling of social safety nets to the violent capture of existing infrastructure \cite{becker_ideology_2020,grynkewich_welfare_2008}. Identifying those psychosocial factors that are common across extremist ideologies — if any — can offer valuable insights into why individuals are drawn to extremist movements, regardless of their ideological differences.

Decades of research have explored the psychological and social aspects of extremism. Approaches have focused on identifying psychological traits using psychometric evaluations~\cite{corner_reviewing_2021} and various belief systems using survey-based methods requiring individuals to assess their level of agreement with statements such as ``Foreigners and asylum seekers are the ruin of (country)" and ``Under some circumstances, a nondemocratic government can be preferable" \cite{jungkunz2024measuring}. However, ethical and logistical barriers to directly engaging extremist and terrorist communities limit the scalability of these research methods and pose challenges for conducting comprehensive empirical studies on the most relevant populations. Meanwhile, the rapid mobilization of extremist movements via social media produces vast amounts of data that is not only a rich and underutilized data source for applying such frameworks but also urgently needs to be investigated in order to develop prevention and intervention strategies. 

The primary objective of this work is to determine whether we can leverage existing, theory-based psychosocial frameworks to analyze social media data and significantly expand empirical insights into extremist belief systems and communities at scale. Specifically, we investigate the following three research questions:

\textbf{RQ1.} Can an unsupervised method combining NLP techniques and psychosocial theories of extremism identify characteristics of ideological extremism in online communities?

\textbf{RQ2.} How do these identified traits provide insights into the psychological composition and ideological similarities or differences across various extremist groups?

\textbf{RQ3.} Can our framework reliably signal the likelihood of an individual's eventual active engagement with an extreme ideological community based on their evolving language patterns?

We aim to explore and understand how various psychological and social components contribute to — and characterize — radicalization across a range of extreme ideological groups.~\footnote{See \S\ref{app:term_def} for semantic clarification.} 
Given that numerous theories have been developed and applied to particular extremist groups and belief systems, we explore a method to unify them in a single model, allowing us to extract factors of extremism in broad, unsupervised data. To make use of existing psychometric and survey-based tools for measuring an individual's tendency toward extremism at a large scale on social media data, we leverage natural language processing (NLP) techniques to obtain scores for the items using only the available text, requiring no manual labeling or supervision.  We identify 11 orthogonal dimensions — the ``Extremist Eleven" — that reflect important factors that can be used to broadly characterize key traits linked to extremist ideologies (\textbf{RQ1}). These dimensions are psychosocially meaningful and offer a more complete understanding of the cognitive and emotional drivers that underlie engagement with extremist groups. We explore the face validity of our method by examining how these factors emerge in communities known for extreme-ideological discourse, and in a diverse representation of social media text that captures political discourse more broadly, as well as non-topic-specific discourse.

Additionally, we demonstrate the utility of our methods and results in two preliminary ways. First, we show how the Extremist Eleven can be applied to characterize the psychological composition of both new and existing online communities, offering insights into the shared and unique traits that define these spaces (\textbf{RQ2}). Second, we find that we can accurately and reliably identify individuals who are likely to affiliate with and bind themselves to an extremist online community well in advance — more than 6 months prior to actively joining such communities (\textbf{RQ3}). This predictive capability highlights the practical application of our approach for early detection and intervention in online radicalization.
The code and implementation details are available to the public 
and data can be made available upon request.

\begin{table*}[ht]
\small
    \centering
    \begin{tabular}{lllr}
    \toprule
    \textbf{Scale} & \textbf{Source} & \textbf{Construct(s) Assessed} & \textbf{\# of Items} \\
    \midrule
      Extremism Scale   &  \citet{ozer-2018-capturing} & General Extremist Attitudes  & 14 \\
       Social Dominance Orientation  & \citet{ho2015nature}& Preferences for Social Hierarchy & 8 \\
       Radicalism Intention  & \citet{moskalenko2009measuring}& Propensity for Radical Action & 4 \\
       Violent Intention  & \citet{obaidi2018living, obaidi2018mistreatment}& Likelihood of Violent Behavior  & 7 \\
       Nationalism Scale   & \citet{weiss2003cross} & National Identity and Superiority  & 4 \\
       Right-Wing Authoritarianism   & \citet{zakrisson2005construction} & Obedience to Authority and Tradition  & 15 \\
       Self-Categorization Scale   & \citet{ellemers1999self} & Group Identity and Affiliation  &3 \\
       Dirty Dozen  & \citet{jonason2010dirty} & ``Dark'' Personality Traits &12 \\
       General Extremist  & \citet{jungkunz2024measuring} & Broad Extremist Tendencies &5 \\
       Left-Wing Radical  & \citet{jungkunz2024measuring} & Radical Left Ideological Views &6 \\
       Right-Wing Radical  & \citet{jungkunz2024measuring} &  Radical Right Ideological Views &7 \\
       Ethnic Intolerance & \citet{weiss2003cross} & Xenophobic Attitudes &  4 \\
       \bottomrule
    \end{tabular}
    \caption{Extremism scales used in our model and keywords describing the relation the scale has to extremism.}
    \label{tab:extremism_scales}
\end{table*}

\section{Expanding the Psychometric Toolkit in the Context of Extremism}

Over the past 80 years, psychologists and social theorists have extensively explored the psychological and social factors related to extremist beliefs and behaviors. One of the earliest and most influential contributions to the field was Theodor Adorno’s ``F-scale,'' developed in the aftermath of World War II to measure authoritarian tendencies in individuals ~\cite{adorno_authoritarian_1950}. The F-scale sought to better quantify an individual's susceptibility to authoritarian attitudes, suggesting that individuals with certain personality traits — such as rigid adherence to conventional values, submission to authority, and aggression toward out-groups — were more likely to adopt extreme ideologies. 

Since then, a range of psychological constructs have been explored in relation to extremism~\cite{corner_reviewing_2021}. The Dark Triad, for example, which comprises narcissism, Machiavellianism, and psychopathy, has been studied as a cluster of personality traits associated with manipulative and antisocial behavior. Research in this area has suggested that individuals scoring high on these traits may be more susceptible to radicalization due to their callousness, desire for power, and disregard for social norms~\cite[e.g.,][]{jones_psychopathy_2013, awan_cyber-extremism_2017}. Similarly, dogmatism — a rigid and inflexible cognitive style that resists contradictory evidence — has been identified as an important corollary to extremist thinking. Dogmatic individuals are more likely to reject alternative viewpoints that contradict their extreme positions~\cite{van_prooijen_extreme_2017}, often seeing the world in black-and-white terms. These domains, among others, have provided valuable insights into the psychological underpinnings of extremist ideologies, illustrating how certain personality traits and belief systems can predispose individuals to radicalization.

However, while these traditional research methods have advanced our understanding of extremism, they present significant limitations when applied to contemporary extremist groups. Accessing individuals from white supremacist, incel, or terrorist organizations poses considerable ethical and logistical challenges, making it difficult to directly study these populations using surveys or laboratory-based methods~\cite{egan_can_2016}. This limitation has often restricted researchers to using small, non-representative samples or retrospective accounts ~\cite[see, e.g.,][]{gaudette2022role}, which can fail to capture the full complexity of these groups’ beliefs and behaviors. As a result, much of the empirical research on extremism has relied heavily on theoretical constructs without the ability to systematically engage with the most active and radicalized individuals.

Recent advances in natural language processing (NLP) have opened up novel possibilities for addressing this gap. By analyzing the language used by extremist communities in online forums, social media, and other digital spaces, researchers can infer the ideological content and psychological dispositions of these individuals. NLP techniques allow for large-scale analysis of the beliefs expressed by individuals in these communities, providing a new method for assessing constructs such as dogmatism, authoritarianism, or the Dark Triad. This approach enables researchers to approximate the results of psychological questionnaires by measuring the degree to which extremist language aligns with established psychological theories. In this way, language can serve as a proxy for direct survey responses, offering new insights into the psychological dimensions of extremist ideologies without the need for direct engagement with these hard-to-reach populations.

\section{Integrating NLP with Social Psychological Theory for Enhanced Extremism Research}

Recent work has demonstrated that combining NLP with psychological theory can yield insights that go beyond what either field could accomplish in isolation. For example, \citet{varadarajan_archetypes_2024} demonstrated the power of contextualized embeddings for detecting suicidality by quantifying user language for ``archetypal'' representations of suicidological theory and constructs (i.e., perceived burdensomeness, thwarted belongingness, and acquired capability). 
Similarly, \citet{atari_contextualized_2023} generated contextualized embeddings of existing, theory-based self-report questionnaires for the explicit purpose of assessing such constructs in naturalistic data, bypassing the requirement for questionnaire-based assessments. 
In the same vein, other work has combined NLP with theories of morality and political communication to study changes in language conveying dehumanizing attitudes surrounding marginalized groups \cite{mendelsohn2020framework} and relationships between ideology communicative frames in immigration discourse \cite{mendelsohn-etal-2021-modeling}. These studies have demonstrated the emerging potential of quantifying psychosocial constructs through natural language data in a manner that \textit{explicitly} derives linguistic representations from theory, providing new avenues for psychometric analysis. By merging rigorous computational methods with well-established psychological frameworks, researchers can enhance the diagnostic value of natural language data, offering a richer understanding of how extremist traits emerge, are expressed, and persist across time and context.

In recent years, NLP and text analysis have become indispensable tools for studying extremism, particularly in the context of online communities. Several computational approaches have been developed to analyze and track extremist behavior online. One notable example is the work by \citet{cohen2014detecting}, who introduced NLP techniques to identify ``warning behaviors'' associated with radical violence. They focused on three key behaviors: leakage (the communication of intent to harm), fixation (obsessive focus on a cause or target), and identification (associating with a militant ideology or persona). Their study provided a framework for detecting these behaviors through linguistic markers, such as verbs expressing intent (e.g., “I will...,” or “someone should...”) and frequent references to out-groups. This approach highlighted how text analysis could reveal early signs of extremist tendencies, especially in online environments where direct observation is often challenging.

Building on this, \citet{hartung-etal-2017-ranking} modeled right-wing extremist behaviors by analyzing Twitter data. Their method involved measuring the similarity between user profiles and known extreme or non-extreme ideological groups, offering a nuanced view of how individuals align with extremist ideologies. Unlike traditional binary classification systems, their approach used a continuous scale, allowing for a more dynamic understanding of how individuals engage with extremist content over time. However, the limitation of relying on hand-crafted features and predefined extremist profiles pointed to the need for more flexible models that could capture the evolution of extremism across diverse groups.

Other studies have focused on specific ideological communities, offering more detailed insights into how language fosters radicalization. For example, \citet{ribeiro2020pick} analyzed the language of incels (involuntary celibates) and other communities in the ``manosphere,'' tracing the pathways through which users transition between more moderate and extreme subgroups. This large-scale analysis of online forums revealed how the use of hostile and misogynistic language serves as a gateway to more radical ideologies. 
Additionally, \citet{dragos-etal-2022-angry} examined the role of emotion in extremist discourse, correlating human annotations of emotion (e.g., anger, hatred) with judgments of whether the content was extremist or not. Their findings underscored the importance of emotional language in radicalizing individuals and justifying extremist actions.

While these computational methods have significantly advanced our ability to detect and analyze extremist behaviors in online spaces, they often operate independently from the substantial body of research in the social sciences that examines the underlying drivers of radicalization. The work being done in NLP typically focuses on the linguistic and behavioral manifestations of extremism, while research in psychology and sociology often seeks to understand the deeper cognitive, emotional, and social processes that make individuals susceptible to extremist ideologies. These two fields, though complementary, stem from distinct intellectual traditions and often pursue different questions \cite{boyd2024verbal}, with computational approaches prioritizing large-scale analysis and detection, and social science approaches aiming to uncover the psychological and social mechanisms that lead to radicalization. Despite their potential for synergy, these areas of research remain somewhat disconnected.

Our work highlights opportunities to bridge these two approaches — computational analysis and psychological theory — to push our understanding of extremism further. While NLP can effectively identify patterns in language and behavior at scale, wedding these methods with psychological theories of the emotional, cognitive, and social vulnerabilities that drive individuals toward extremism allows for a more holistic approach. This interdisciplinary synthesis enables the development of models that not only detect who is participating in extremist discourse but also explain why individuals are drawn to these movements, offering deeper insights into the underlying mechanisms of radicalization.

\begin{figure*}[htb]
    \centering
    \includegraphics[width=0.95\linewidth]{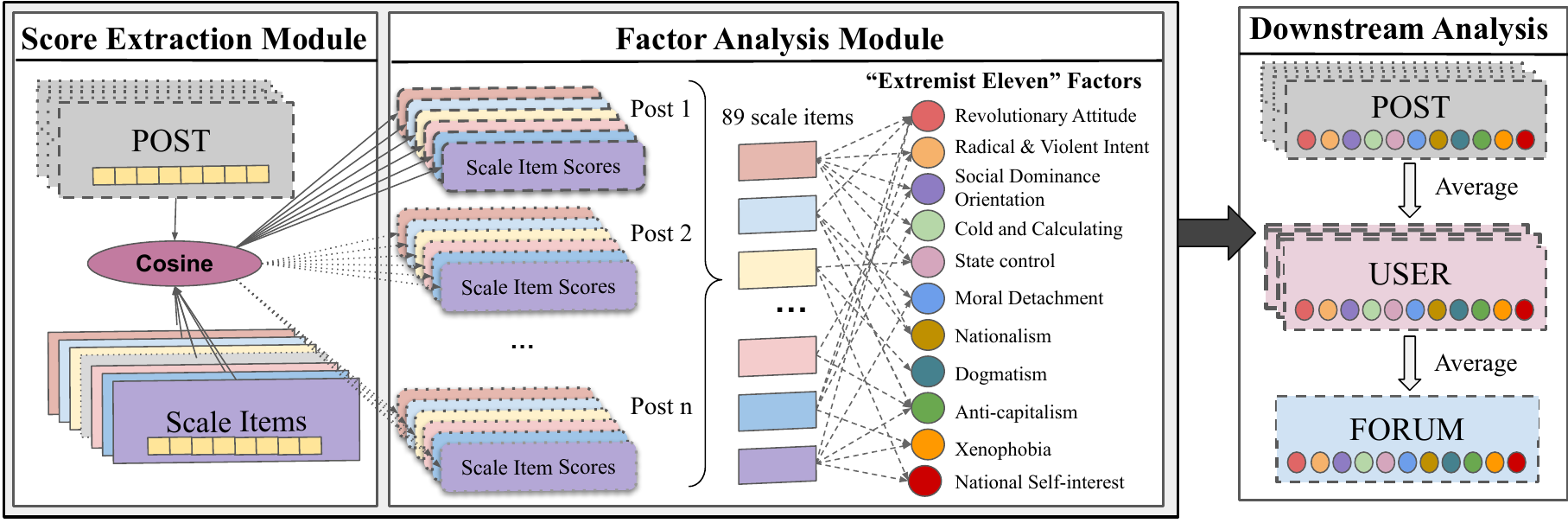}
    \caption{
    Flowchart of our method. We extract the Scale Scores using cosine similarity of the sentence representations of each post against the 89 scale items. Then, these scores are treated as item scores for a Factor Analysis module, which yields 11 distinct factors that we call ``Extremist Eleven''. While these 11 Extremist Eleven scores are extracted for each post, we conduct large-scale analysis by aggregating them at user- and forum-level.
    }
    \label{fig:method_flowchart}
\end{figure*}

\section{Data}
\label{sec:data}
We study the unifying characteristics of extremism across various ideological groups, especially those that have been associated with real-world violent attacks. We include (1) posts from the white supremacist/neo-Nazi forum \textit{Stormfront}; (2) posts from incel communities on Reddit; and (3) quotes from religious or political leaders published in ISIS periodicals (i.e., Dabiq and Rubiyah). To distinguish between general political discourse and extremist discourse, we include a subset of subreddits from the Politosphere dataset -- including multiple subreddits that have since been banned. Further, to contrast the general rhetoric on social platforms and how they compare against the extremist discourse online, we also collected general posts from a set of users on Reddit selected at random. We removed non-English posts, posts with fewer than 10 words, and posts before 2010-01-01 for a more consistent time range since much of the community was not as active and did not exist before then.

\subsection*{Narratives of known extremist discourse}

\noindent \textbf{1. White Supremacist Forum (WS)} We collect a subset of posts from the Stormfront dataset introduced by~\citet{van2021grievance}. Stormfront is a white supremacist web forum that began in the 1990s and has been documented to propagate ideas of the radical right, including neo-Nazism and white nationalism~\cite{bowman2009exploring, caren2012social}. This dataset contains 1,782,499 posts by 52,203 users from 2001-09-11 to 2015-02-01.  We included users that had posted at least ten times in that time period. We included the most posted forums within the dataset: \textit{Politics \& Continuing Crises, Lounge, Ideology and Philosophy, For Stormfront Ladies Only, Strategy and Tactics, Talk}. After all filters, this set has 173,359 posts from 3,622 users. The average post length for this dataset was 
$132\pm246$ words, with an interquartile range (IQR) of 48--138 words. 

\textbf{2. Incel Reddit (IR)} Incels participate in online communities where misogyny and calls for violence against women are prevalent, fueled by the belief that women are withholding a perceived ``right'' to sex. We collected comments and posts from three subreddits: r/Incels, r/Braincels, and r/Trufemcels, with posts ranging from 2016-07-01 to 2020-04-30. This timeline includes the period when r/Incels was banned (November 2017) for violating Reddit's policy against incitement of violence toward women, and the subsequent creation of r/Braincels, which gained increasing popularity throughout 2018. The availability and volume of data also allowed us to examine users who began interacting with these forums during this period (\S\ref{sec:user-level}). In total, we collected 51,266 posts from 1,271 users (34,259 posts on r/Braincels, 14,178 posts on r/Incels, and 2,829 posts on r/Trufemcels), with an average post length of $109\pm160$ words. The IQR was 30-129 words. These three forums were chosen for their high volume of posts and to facilitate a comparison between incels and femcels, who use similar extremist narratives despite being on opposing sides of the ideological spectrum. 

\textbf{3. ISIS Articles} Given limited accessibility of English-speaking Islamic extremism discourse on internet forums, we use a dataset consisting of religious and ideological quotes from two ISIS-supporting magazines: \textit{Dabiq} and \textit{Rumiyah}, used to promote their propaganda~\cite{ISISdataset}. It includes scraped data from 15 issues of \textit{Dabiq} (6/2014 to 7/2016) and 9 issues of \textit{Rumiyah} (9/2016 to 5/2017), resulting in 2,685 texts. The average message length was 
$70\pm127$ and the IQR is 22--71 words. Since the dataset consists solely of quotes from religious and political leaders, no user-level analysis was conducted. Instead, it was used exclusively to learn relevant factors related to Islamic extremism.

\subsection*{General discourse}

\noindent \textbf{1. General Reddit (GR)} We selected 100 random users who posted or commented at least five times in October 2015 and collected their posts from 2014-01-01 to 2017-01-01. We only retained posts from users who were active for at least 20 out of the 24 months, ensuring that we had users who consistently engaged on Reddit over a couple of years. This process resulted in a dataset covering a wide range of topics, with the five most commonly posted subreddits being: \textit{movies, ProtectAndServe, nfl, loseit,} and \textit{AskReddit}, totaling 63,666 posts. The average post length was 
$47\pm62$ words. The IQR was 18-53 words.

\textbf{2. Politosphere Reddit (PR)} We collected comments and posts from the Politosphere Reddit dataset \cite{hofmann2022reddit}, which consists of subreddits focused on political discourse. This dataset provides us with two unique analysis opportunities. First, many of the items on the extremism scales are inherently or explicitly political in nature, as shown in Table \ref{tab:extremism_scales}. Since extremism is often studied in politico-religious contexts, and both extremist and general political discourse may share similar lexical features, this dataset provides a valuable point of comparison. Second, the meta-data of Politosphere indicates whether the subreddit has been banned, which allows us to explore whether and how extremist factors are present in communities that have violated Reddit policies. Therefore, we sampled subreddits that have been banned as well as those that have not, to explore whether banned subreddits exhibit extremist characteristics. From the collection of 605 subreddits, we randomly selected nine banned and nine not-banned subreddits (see \S\ref{app:subreddits}) resulting in 591,066 posts, with an average post length of $78\pm113$ words. The IQR is 22-87 words.

\section{Extremism Scales}

Our method expands the reach and applicability of existing psychological measures designed to quantify traits linked to extremism~\cite{obaidi2022measuring}. Some of these scales were developed to measure specific extremist ideologies, while others focus on more broadly harmful social traits. In this work, we combine all relevant indicators from each scale to create a unified model of extremism factors. In total, we applied 89 items from 10 distinct measurement scales (see Table~\ref{tab:extremism_scales}). Our goal was to develop a broadband framework that encompasses a wide range of constructs, spanning past research across various frameworks ranging from general models of extremism to specific ideologies, personality traits, and factors like nationalism and ethnic intolerance. Since these indicators originate from direct self-report questionnaires, all items are phrased as first-person statements, allowing individuals to rate their level of agreement.

\section{Method}\label{sec:method}

Our method is comprised of a Scale Score Extraction Module and a Factor Analysis Module illustrated in Figure~\ref{fig:method_flowchart}.

\subsubsection{Scale Score Extraction.}
To score the language content based on the 89 extremism scale items, we employed a method that computes cosine similarity between the vector representations of the post content and each extremism item. This technique, variously referred to as ``archetypes'' \cite{varadarajan_archetypes_2024} or contextualized construct representation \cite{atari_contextualized_2023}, allows us to quantify the alignment of posts with extremism-related traits. We encode both the posts and survey items using the \texttt{mxbai-embed-large-v1} model from MixedBread \cite{emb2024mxbai, li2023angle} using the default hyperparameters. To control for variations in post length, we split each post into 100-word chunks, reducing the impact of length on similarity across our cross-domain datasets~(\S\ref{sec:data}). The chunk-level scores are then mean-aggregated to produce post-level extremism item scores.

\subsubsection{Exploratory Factor Analysis-based Scoring.}
Questionnaire scale items are typically factor-analyzed to identify distinct, interpretable factors representing the latent constructs being measured~\cite{shrestha2021factor}. We apply exploratory factor analysis (EFA) to the 89 extremism scale items pooled together across all the datasets combined: White Supremacist Forums, Incel Reddit, ISIS Articles, General Reddit, and Politosphere Reddit, resulting in 882,042 data points.

We conduct three key assessments of the suitability of our analysis. First, the Kaiser–Meyer–Olkin (KMO) measure~\cite{kaiser1974educational} was used to evaluate sampling adequacy based on the correlation matrix, yielding a KMO value of 0.926, indicating strong suitability for factor analysis. Second, Bartlett’s Test of Sphericity~\cite{gorsuch1973using} was used to assess whether the correlation matrix differs significantly from an identity matrix, with a resulting $p$-value $<$ .001, confirming the appropriateness of the dataset for EFA. Finally, we used Horn’s parallel analysis~\cite{horn1965rationale} to determine the optimal number of factors for EFA, which was revealed to be 11 factors.

\section{Results and Discussion}
\label{sec:results}

\subsection{Discovered Factors}

\begin{table}[t!]
    \centering
    \small
    \begin{tabular}  {|l|}
    \hline
    \multicolumn{1}{|c|}{\textsc{National Self-Interest}} \\
    \hline
         \noindent\parbox[c]{\linewidth-0.5cm}{{\textbf{White Supremacist Forum}\\``Exactly! \hlc[blue!25]{We do not gain anything by supporting [COUNTRY] ourselves.} It would be much better for everyone if [COUNTRY] and [COUNTRY], \hlc[blue!25]{devoid of outside assistance or interference, both slaughtered each other}''}}  \\
         \noindent\parbox[c]{\linewidth-0.5cm}{{``\hlc[blue!25] {I am strongly against ANY foreign aid(except perhaps White nations)}, but I believe that giving aid to [COUNTRY] only is even worse than the debt created by giving aid to all ME countries.'' }}  \\ 
         \hline
         \noindent\parbox[t]{\linewidth-0.5cm}
         {\textbf{General Reddit}\\``They're moving production of Oreos to [COUNTRY] to cut costs anyway. No thanks, \hlc[blue!25]{I'd rather support American jobs.}'' }  \\
         \noindent\parbox[t]{\linewidth-0.5cm}{``the trade agreement was a bad deal.  \hlc[blue!25]{How do you think the Average American benefited from this arrangement?}'' }  \\
         \hline       
    \end{tabular}
    \caption{Characteristic differences in posts that signal National Self-Interest in ideological versus general groups.}
    \label{tab:national_self_interest}

\end{table}
\begin{table}
    \centering
    \footnotesize
    \begin{tabular}  {|l|}
    \hline
    \multicolumn{1}{|c|}{\textsc{Cold and Calculating}} \\ 
    \hline
         \noindent\parbox[t]{\linewidth-0.5cm}{{\textbf{Incel Forum}\\I know it's unhealthy, but I can't stop thinking about how much I'd like to prove myself to others- even though I know it's pointless. \color{black}\hlc[blue!25]{I'd just like to show everyone how much of a morally-superior and better person I am overall, just to spite them. }}}  \\
         \hline
         \noindent\parbox[t]{\linewidth-0.5cm}{\textbf{White Supremacist Forum}\\Often I am flying under my own radar. \hlc[blue!25]{I can know and understand the social rules and still violate in an obtuse way that I can't see....Until someone is pissed! But usually I'm just having fun pushing buttons until I find the right one, or enough of the wrong ones.} I tend to make people think for some time after an interaction.}  \\
         \hline
         \noindent\parbox[t]{\linewidth-0.5cm}{\textbf{Femcel}\\If I ever get a people-people job, \hlc[red!25]{I am going to make sure I treat below-average people like me even better because they don't even get 1/12th of the love that the rest of society gets}.} \\
         \hline
    \end{tabular}
    \caption{Strengths and Limitations: Examples where our method successfully identifies extremist traits (blue) and one where it finds a false positive (red). } 
    \label{tab:factor_examples}
\end{table}

\begin{figure*}[th!]
    \centering
    \includegraphics[width=\linewidth]{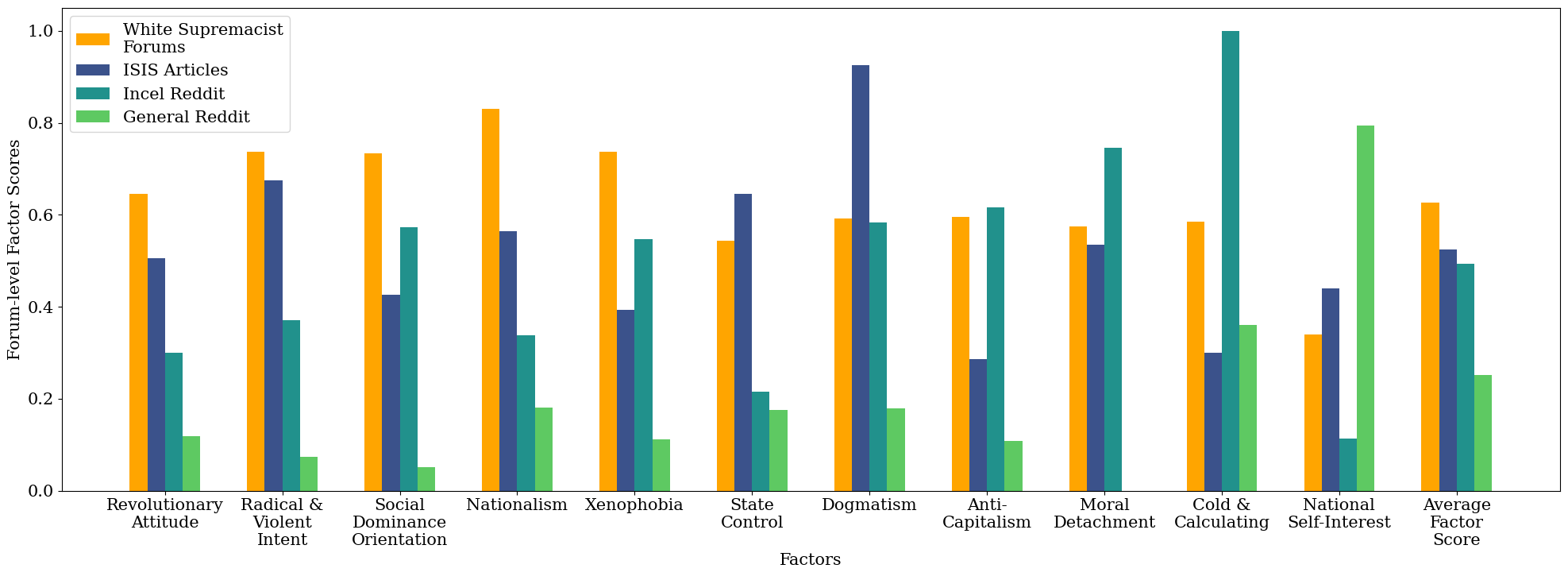}

    \caption{
    Average factor scores for each forum. Higher factor scores suggest the presence of a given extremist trait, while lower scores indicate the absence of such traits. A positive average factor score may signal that a forum is likely to have users who exhibit extremist behaviors, whereas negative scores suggest that extremist tendencies are minimal.
    }
    \label{fig:forum_factor_scores}
\end{figure*}

Starting with \textbf{RQ1}, our goal was to examine how different psychological and social factors contribute to and define radicalization across various extremist groups. We developed an unsupervised model to analyze extremism in online media by comparing the language in posts to established psychological survey scales, followed by factor analysis to score the results. This approach led to the identification of 11 distinct dimensions, which we refer to as ``The Extremist Eleven.'' Table~\ref{tab:extremist_eleven} lists the names of these dimensions, derived from analyzing the content most strongly linked to each factor. These dimensions capture key psychological and social traits commonly associated with extremist ideologies, such as revolutionary attitudes and tendencies toward radical and violent actions.

To assess the face validity of our model, we compared how these factors manifest in extremist communities (White Supremacist, Incel Reddit, and ISIS Articles) versus the broader General Reddit discourse. Figure~\ref{fig:forum_factor_scores} shows the average factor scores for each group. Positive bars indicate that the factor is a prominent feature of that group, while negative bars indicate its absence. At first glance, the model demonstrates strong face validity. For example, violent groups like White Supremacists (WS) and ISIS score higher on factors such as \textit{revolutionary attitudes} and \textit{radical or violent intentions}. We also observe distinctions between WS and ISIS, with the former scoring higher on \textit{Social Dominance Orientation} and \textit{Nationalism}, and the latter on \textit{State Control} and \textit{Dogmatism}. In contrast, General Reddit shows no significant presence of these extremist factors, except for \textit{National Self-Interest}. Upon closer examination, this could be explained by the proportionally frequent discussions on news and current affairs in General Reddit, as opposed to the more extreme rhetoric found in extremist groups. This factor appears in 60\% of General Reddit posts, compared to 38\% in White Supremacist posts. Table~\ref{tab:national_self_interest} provides examples of the highest-scoring posts for this factor, illustrating how extreme discourse, characteristic of White Supremacist forums, is absent in the top General Reddit posts.

Table~\ref{tab:factor_examples} presents both positive and negative examples of our model's effectiveness. In the \textit{r/Femcel} example, the model assigns a high score on the ``cold and calculating'' factor, though the trait exhibited may not fully align with this factor. This mismatch may occur due to strong lexical and syntactic similarity, rather than genuine content similarity, with the survey items. In early experiments, we explored techniques like NLI and dissonance detection to better capture agreement with the statements. However, cosine similarity proved to be an effective compromise, offering solid performance while balancing computational efficiency. Future work could enhance this approach by employing an ensemble of models tailored to the specific nuances of the survey items, potentially improving accuracy in such edge cases.

\subsection{Forum-level Characterization}

We analyze the emergence of the Extremist Eleven factors in communities representing different extremist ideologies to identify whether certain psychosocial factors are common across ideologies and whether they reveal characteristics of different forms of extremism (\textbf{RQ2}). 

 \begin{figure*}[!th]
     \centering
     \includegraphics[width=0.9\linewidth]{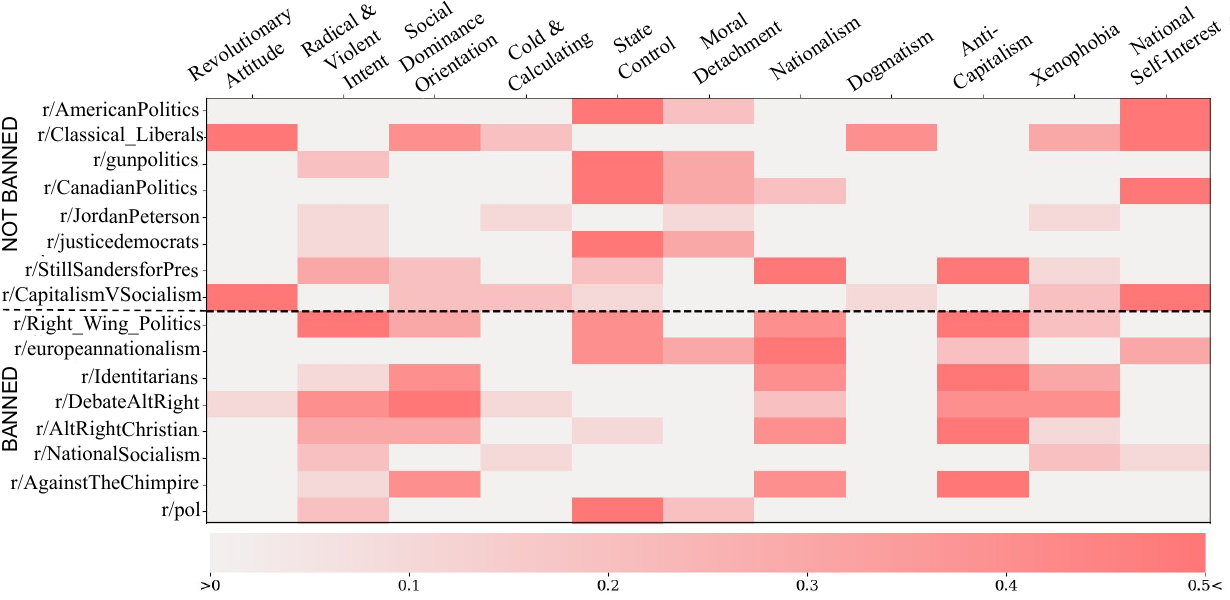}
     \caption{ Based on the aforementioned heuristic, we focus only on factor scores $>$ 0. The Politosphere dataset includes a wide range of political forums, often engaging in debates and diverse viewpoints. While many of these forums involve standard political discourse, some show signs of developing extremist tendencies, as observed here. Notably, banned subreddits show significantly elevated scores on dimensions related to radical and violent intent. 
     }
     \label{fig:politosphere_analysis}
 \end{figure*}

We observe different flavors of extremism based on those factors associated with different forums. As observed in Figure~\ref{fig:forum_factor_scores}, ISIS scores relatively higher in state control and dogmatism, factors that relate to the enforcement of ideology-rooted morals, whereas white supremacist forums have relatively higher signals of social dominance, nationalism, and xenophobia, which could align with Alt-right sentiments. 
 Figure~\ref{fig:politosphere_analysis} illustrates the factors in specific the political discourse subreddits in Politosphere. 
Meanwhile, Incel groups exhibit a distinctive psychological profile compared to other extremist groups, marked by higher levels of moral detachment and cold, calculating interpersonal traits. This suggests that, unlike groups driven by ideological zeal or collective identity, incels tend to view social interactions and moral considerations in an emotionally detached, utilitarian way. They are more likely to rationalize harmful attitudes and behaviors without guilt or empathy, framing their grievances in personal rather than overtly political or ideological terms. While other extremist groups may be fueled by a sense of moral superiority or ideological righteousness, incels appear to exhibit a more self-serving and emotionally disengaged approach~\cite[see, e.g.,][]{wiggins_conceptions_1989}, focusing on personal victimhood and perceived social injustices without regard for broader ethical consequences. This distinct profile positions incels as less ideologically driven but more inwardly focused on their own grievances and frustrations, which they justify through a detached, often hostile worldview.

As described in \S\ref{sec:data}, it is important to note that the ISIS dataset presents a qualitatively different type of data compared to the others. While the other datasets consist of personal discourse from social media forums, the ISIS data originates from magazine publications, reflecting a distinct, formal style. Despite this difference in format and style, our model demonstrates strong generalization across these varied domains, illustrating the method's effectiveness beyond just personal discourse.

Next, we examine how our model functions within a broader corpus of political discourse without distinguishing extremist-leaning communities, for identifying areas that may indicate warning signals of extremist discourse. In Figure~\ref{fig:politosphere_analysis}, we analyze the factors in samples from both banned and non-banned subreddits within Politosphere. A key preliminary observation is that there is a stronger signal for Radical \& Violent Intent and Xenophobia among the banned subreddits. We performed Student's t-tests on the factor distributions between the banned and non-banned subreddits, confirming that the means are significantly different across all factors. Future work could analyze how the factors change over time, leading up to the point at which they were banned. 

\subsection{User-level Analysis}
\label{sec:user-level}

To address \textbf{RQ3}, we tested the utility of our approach and model by investigating our ability to predict an individual's likelihood of joining an extremist community based on their language patterns in online discourse. Specifically, we applied our method to users who later joined the Incel Reddit community as a case study for radicalization. Morever, we explored its ability to reveal temporal changes in extremist traits leading up to, and following, a person's decision to actively join the Incel Reddit community.

\noindent We further analyze two applications of our method to understand the evolution of extremism at a user level in the context of online communities, where joining or engaging for the first time with an extremist community can be a proxy for radicalization in real life. We compare the Extremist-Eleven Scores of 100 General Reddit users who newly engaged with a forum at some point, and  82 Incel users who were engaged for at least 10 out of 12 consecutive months in the incel community.
\subsubsection{Forecasting active engagement with an extremist community}
\begin{figure}[!htb]
    \centering
    \includegraphics[width=0.9\linewidth]{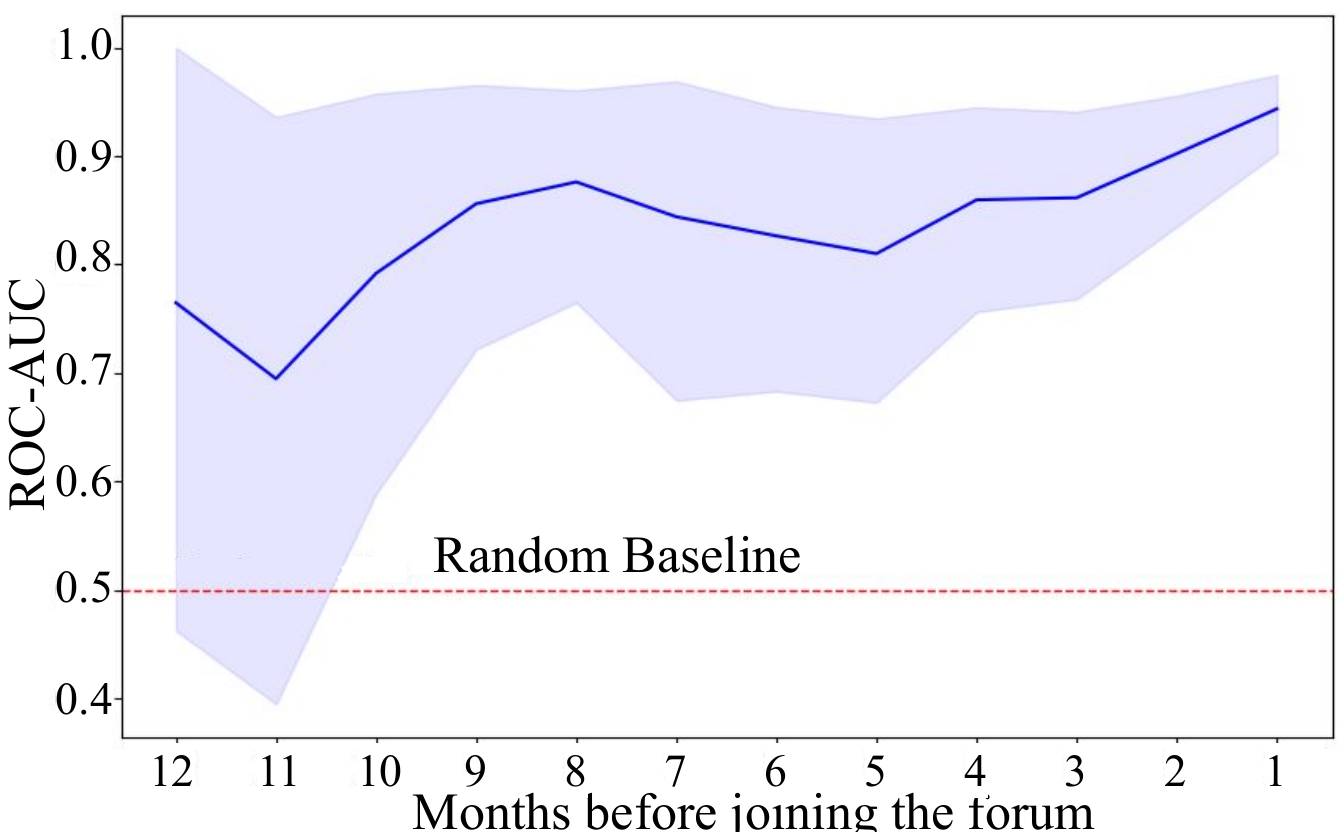}
    \caption{Early detection power of the Extremist Eleven scores for identifying whether users will actively engage with the incel community in the future. The shaded region represents the 95\% confidence interval of the predicted AUC scores. We find that the prediction is already better than chance 8-10 months before a user joins the incel forum.}
    \label{fig:join-community}
\end{figure}

Figure~\ref{fig:join-community} shows the performance of a logistic regression model on the average Extremist-Eleven scores derived from posts up until a certain point in time to predict joining the incel forum in the future. To this end, we collected the history of posts made by the 82 incel users in the 12 months leading up to joining the forum.
For the general users, we use the time when they post on any new forum (except the incel community) as the time of joining. We perform 5-fold stratified cross-validation, reporting the averages of the ROC-AUC across the five runs. 
Figure~\ref{fig:join-community} demonstrates that individuals who would later go on to engage with the incel community exhibited higher scores on extremist traits well before joining. These elevated scores, up to 12 months prior to joining, reliably differentiate users who are at risk of radicalization long before they actively affiliate with such communities with increasingly high accuracy. This result aligns with theoretical frameworks, such as the ~\citet{adorno_authoritarian_1950}
work on authoritarianism, which aimed to identify individuals susceptible to extremist ideologies rather than those who were already fully radicalized. 
Our findings support the notion that personality traits — in this case, for example, moral disengagement and being interpersonally cold and calculating — are reflected in language use, making it possible to predict who might be drawn into these groups well in advance of actual participation.

\subsubsection{Do extremist traits increase/amplify after joining?}

\begin{figure}[t]
    \centering
    \includegraphics[width=0.9\linewidth]{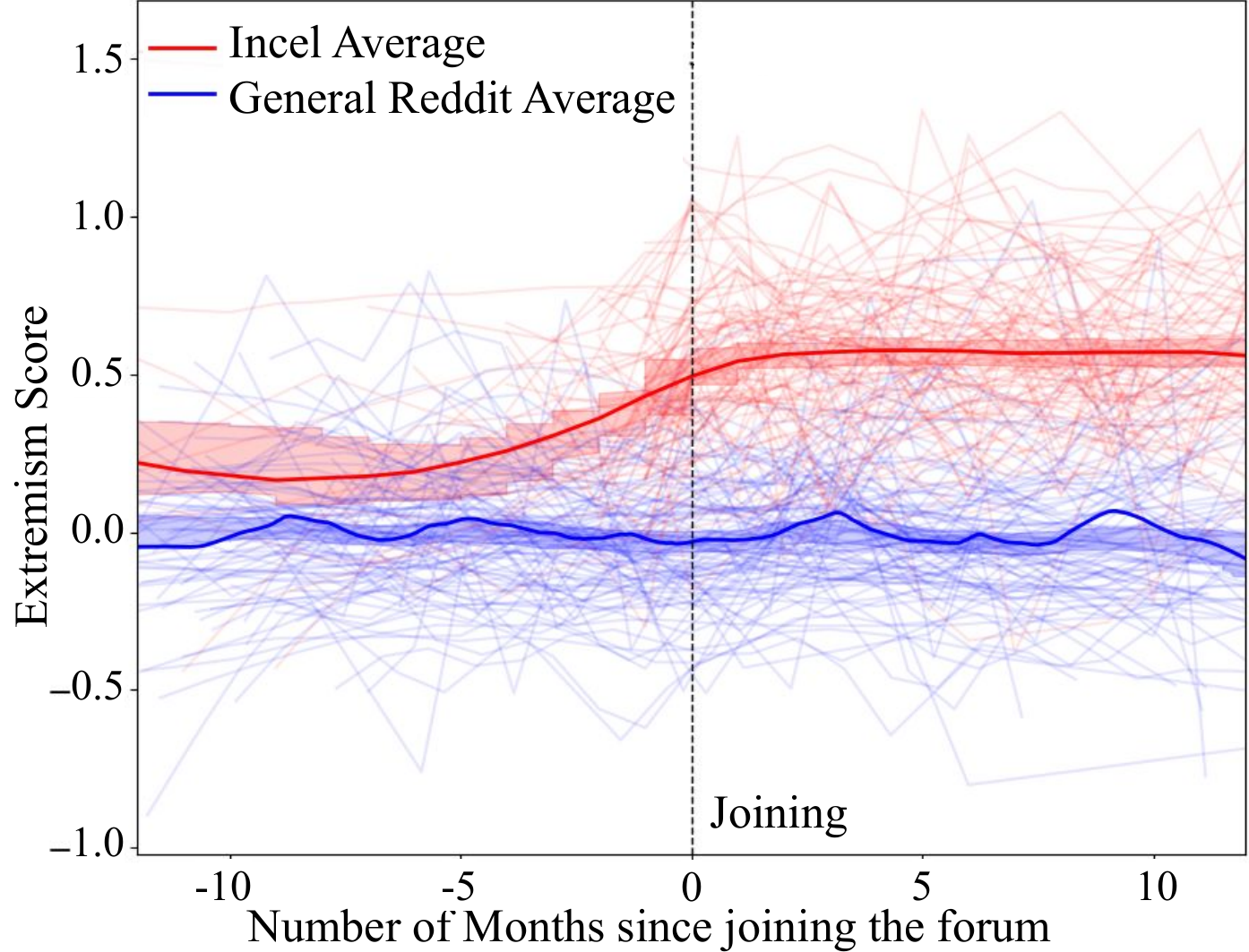}
    \caption{Extremism scores over time for users joining the incel community (red line) and users from General Reddit (blue line). The red line shows an increase in extremist traits leading up to T0, the point at which individuals join the incel community, followed by a plateau 
    post-entry. In contrast, the blue line reflects the extremism scores for General Reddit users, with T0 randomly assigned for each user, and shows no significant change over time. This suggests that while incel members exhibit increasing extremist tendencies before joining, these tendencies stabilize afterward, with no evidence of a similar pattern in the general population.}
    \label{fig:tendency_increase}
\end{figure}

\begin{figure*}[htb]
    \centering
     \includegraphics[width=0.9\linewidth]{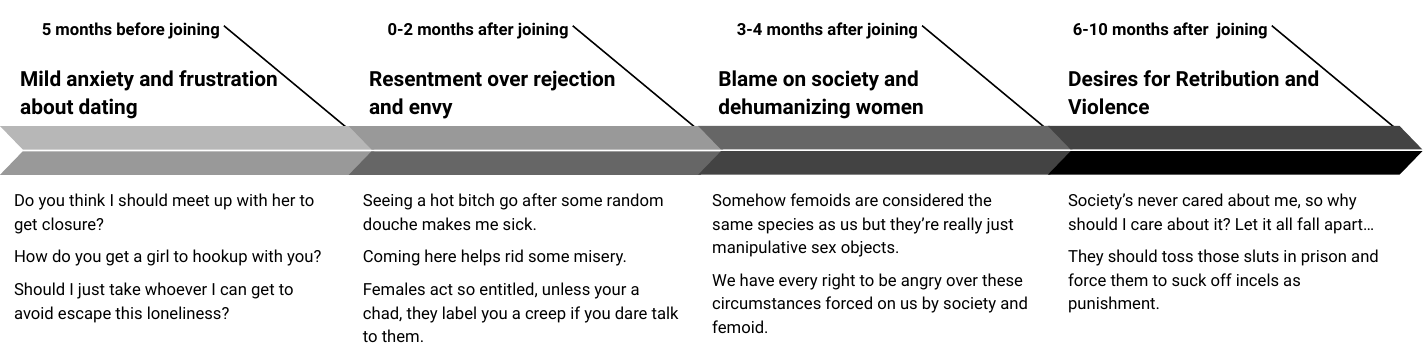}
    \caption{
    \textbf{Incel User Case Study:} A timeline illustrating the phases of an incel community member exhibiting changes in beliefs towards extremism. Five months prior to joining the community, the user exhibited mild anxiety and frustration with dating. Shortly after joining, they show signs of resentment over rejection and envy, while implying a perceived benefit of joining (to help \textit{rid some misery}). At 3-4 months, they begin to blame their circumstances on society and use increasingly dehumanizing language about women; note the progression from \textit{girl}, \textit{females}, to \textit{femoids}. At 6-10 months, the user's rhetoric becomes retributive and violent. The spans are paraphrased extracts from the user's posts.
    }
    \label{fig:user_case_study}
\end{figure*}

Using the same dataset, we additionally examined whether engaging in such communities amplifies extremist traits. We perform local polynomial regression (LOESS) on the extremism scores for each of the users joining the incel community over a period and the users joining a random community over a period of 24 months to derive smooth trajectories of extremism scores.  Figure~\ref{fig:tendency_increase} compares the extremism scores of users in incel forums to those in general Reddit forums, analyzing changes in scores before and after joining a new group. For users joining a random, non-extremist forum, we found no significant change in their extremist trait scores. However, for those joining the incel community, we observed a noticeable increase in these traits leading up to and following their entry into the forum; Fig.~\ref{fig:user_case_study} analyzes the progression of one user's extremism over this timeline. Despite this, the increase was not as dramatic as one might expect, which aligns with recent research questioning the validity of the ``echo chamber'' metaphor in explaining polarization and radicalization~\cite[e.g.,][]{tornberg_echo_2024}. This suggests that joining an extremist group may not automatically intensify extremist tendencies, challenging the assumption that such communities solely function as amplifiers of radicalization.

\section{Conclusion}
\label{sec:conclusions}

In light of the growing influence of extremist groups and their potential for harm, there is a critical need to better understand the mechanisms that drive radicalization. By exploring the psychosocial and ideological dimensions of these movements, researchers can contribute to the development of more effective interventions aimed at reducing their appeal and limiting their growth~\cite{stephens_preventing_2021}.

This study leveraged NLP techniques to extend the applicability of existing psychological measures designed to quantify traits linked to extremism in large-scale data reflecting hard-to-reach extremist populations. We identified the ``Extremist Eleven'' psychological and social dimensions that characterize and distinguish online extremist discourse from general social media discourse. By analyzing these factors, we uncovered commonalities and differences in what characterizes different extremist groups. While the white supremacist and ISIS groups exhibited political, ideological, and violent rhetoric markers, the incel (involuntary celibate) group had a distinct profile that signaled a more inward focus on personal grievances and frustrations. In a case study on active users on incel forums, we found that these individuals exhibited higher scores on extremist traits well before they actively engaged with these forums. Thus, our model may provide insight into who may be susceptible to radicalization. Finally, we observed that extremist traits increased slightly among these individuals after joining incel forums.

The fusion of computational approaches with psychosocial theories represents a significant advancement in the study of extremism, allowing us not only to assess the psychological dimensions of hard-to-reach populations but also to begin exploring the core dimensions that may underpin extremist beliefs and behaviors. 
By combining NLP's capacity for extensive data analysis with psychological frameworks' understanding of human behavior and thought processes, researchers can now conduct more comprehensive studies of extremism. This opens new pathways for investigating radicalization, enhancing our capacity to comprehend and potentially address the psychological mechanisms that draw individuals into extremist groups.

\section{Limitations and Future Work}
\label{sec:limit}
While our unified model of extremist traits provides a novel framework for detecting and predicting radicalization, several limitations exist. 
First, the model is based on existing data and theories, which may not fully capture the complexity and diversity of radicalization pathways across different contexts, cultures, and ideologies. 
The generalizability of the model may therefore be constrained in settings with limited or biased data sources. 
Second, the reliance on certain  psychological traits as predictors may oversimplify the radicalization process, potentially overlooking the influence of situational, social, or political factors. 
Additionally, the model’s predictive accuracy may be affected by inherent biases in the data, including the underrepresentation of certain groups or the risk of overfitting to specific types of extremism. 
Finally, ethical considerations regarding the use of this model for early detection raise concerns about privacy, the potential for false positives, and the stigmatization of individuals based on predictive analysis. Further research is required to address these limitations and to refine the model for more robust and context-sensitive applications.

The current work provides opportunities for extending and refining our unified model of extremist traits. First, expanding the dataset to include more diverse and cross-cultural cases of radicalization will be essential for improving the model's generalizability, particularly given the disagreement about the universality of various psychological and social traits~\cite{dong_are_2020}. Incorporating non-traditional data sources, such as social media activity, network analyses, and various other digital traces could enhance the model's ability to accurately detect early signs of radicalization across different environments and ideological groups~\cite[][]{boyd_personality_2020}. While the current framework emphasizes individual traits, radicalization is often influenced by broader social, political, and economic events. The continued development of ``interactionist'' models that combine individual psychological characteristics with situational and sociocultural triggers (e.g., sociopolitical instability) could yield more nuanced and comprehensive insights.

 \section*{Acknowledgements}

We would like to express our gratitude to the researchers and scholars whose work has laid the foundation for this study. Their thoughtful and diligent efforts in exploring the complexities of extremism, social psychology, and computational analysis have been invaluable. We would like to thank Isabelle van der Vegt for providing the Stormfront data and making it available to the research team.

This work was supported in part by a grant from the NIH-NIAAA (R01 AA028032) and a DARPA Young Faculty Award grant \#W911NF-20-1-0306 awarded to H. Andrew Schwartz at Stony Brook University.
The conclusions contained herein are those of the
authors and should not be interpreted as representing the official policies, either expressed or implied, of DARPA, NIH, any other
government organization, or the U.S. Government. 
Lucie Flek and Allison Lahnala were supported by the German Federal Ministry of Education and Research (BMBF) as a part of the AI Research Group program under the reference 01-S20060, by the state of North Rhine-Westphalia as part of the Lamarr Institute for Machine Learning and Artificial Intelligence and by the Bonn-Aachen International Center for Information Technology (b-it) supporting the visiting research stay at Stony Brook University.

\bibliography{custom,ryanboyd}

\section*{Ethics Checklist}
\begin{enumerate}

\item For most authors...
\begin{enumerate}
    \item  Would answering this research question advance science without violating social contracts, such as violating privacy norms, perpetuating unfair profiling, exacerbating the socio-economic divide, or implying disrespect to societies or cultures?
    \answerYes{Yes. Our research contributes toward understanding factors of extremism without needing to engage directly with individuals who may or may not be involved in extremist communities, or have any awareness of their identity. }
  \item Do your main claims in the abstract and introduction accurately reflect the paper's contributions and scope?
    \answerYes{Yes, we have outlined the contributions and scope of our paper clearly.}
   \item Do you clarify how the proposed methodological approach is appropriate for the claims made? 
    \answerYes{Yes, we discuss that in \S\ref{sec:results}.}
   \item Do you clarify what are possible artifacts in the data used, given population-specific distributions?
    \answerYes{Yes, we clarify that lexical and syntactic similarities may be present in both extremist communities and general political discourse.}
  \item Did you describe the limitations of your work?
    \answerYes{Yes, we describe it in \S\ref{sec:limit}}.
  \item Did you discuss any potential negative societal impacts of your work?
    \answerYes{Yes, we describe it in \S\ref{sec:limit}}.
      \item Did you discuss any potential misuse of your work?
    \answerYes{Yes, we describe it in \S\ref{sec:limit}}.
    \item Did you describe steps taken to prevent or mitigate potential negative outcomes of the research, such as data and model documentation, data anonymization, responsible release, access control, and the reproducibility of findings?
    \answerYes{Yes, we address that in \S\ref{sec:ethics}.}
  \item Have you read the ethics review guidelines and ensured that your paper conforms to them?
    \answerYes{Yes.}
\end{enumerate}

\item Additionally, if your study involves hypotheses testing...
\begin{enumerate}
  \item Did you clearly state the assumptions underlying all theoretical results?
    \answerNA{NA, our work was a product of exploration rather than testing hypotheses.}
  \item Have you provided justifications for all theoretical results?
    \answerNA{NA, yet our findings illuminate further avenues of investigation to embed our empirical results into theoretical constructs.}
  \item Did you discuss competing hypotheses or theories that might challenge or complement your theoretical results?
    \answerNA{NA. However, we mention potential pitfalls of our approach in \S\ref{sec:limit}.}
  \item Have you considered alternative mechanisms or explanations that might account for the same outcomes observed in your study?
    \answerNA{NA. We do not make conclusive explanations about the underlying mechanisms.}
  \item Did you address potential biases or limitations in your theoretical framework?
    \answerNA{NA. However, we discuss limitations in \S\ref{sec:limit}.}
  \item Have you related your theoretical results to the existing literature in social science?
    \answerNA{NA. Our work is heavily based off social scientific theories of extremism.}
  \item Did you discuss the implications of your theoretical results for policy, practice, or further research in the social science domain?
    \answerYes{Yes. We discuss potential implications in \S\ref{sec:conclusions}.}
\end{enumerate}

\item Additionally, if you are including theoretical proofs...
\begin{enumerate}
  \item Did you state the full set of assumptions of all theoretical results?
    \answerNA{NA. We do not include theoretical proofs.}
	\item Did you include complete proofs of all theoretical results?
    \answerNA{NA. We do not include theoretical proofs.}
\end{enumerate}

\item Additionally, if you ran machine learning experiments...
\begin{enumerate}
  \item Did you include the code, data, and instructions needed to reproduce the main experimental results (either in the supplemental material or as a URL)?
    \answerYes{Yes, we provide a url to the code on the first page.}
  \item Did you specify all the training details (e.g., data splits, hyperparameters, how they were chosen)?
    \answerYes{Yes. We stated in \S\ref{sec:method} that we use default hyperparameters for the encoder, and discussed the details of the logistic regression model in S\ref{sec:user-level}. }
     \item Did you report error bars (e.g., with respect to the random seed after running experiments multiple times)?
    \answerNA{NA. We do not run experiments multiple times with random seeds.}
	\item Did you include the total amount of compute and the type of resources used (e.g., type of GPUs, internal cluster, or cloud provider)?
    \answerYes{Yes. We state this in \S\ref{sec:compute}.}
     \item Do you justify how the proposed evaluation is sufficient and appropriate to the claims made? 
    \answerNA{NA. Our work is more exploratory than evaluative.}
     \item Do you discuss what is ``the cost`` of misclassification and fault (in)tolerance?
    \answerYes{Yes, we discuss this in \S\ref{sec:ethics}.}
  
\end{enumerate}

\item Additionally, if you are using existing assets (e.g., code, data, models) or curating/releasing new assets, \textbf{without compromising anonymity}...
\begin{enumerate}
  \item If your work uses existing assets, did you cite the creators?
    \answerYes{Yes. See \S\ref{sec:data}.}
  \item Did you mention the license of the assets?
    \answerNA{NA.}
  \item Did you include any new assets in the supplemental material or as a URL?
    \answerNA{NA.}
  \item Did you discuss whether and how consent was obtained from people whose data you're using/curating?
    \answerYes{Yes. We state in \S\ref{sec:ethics} that we collected publicly available data, ensuring no private or personally identifiable data is accessed or used.}
  \item Did you discuss whether the data you are using/curating contains personally identifiable information or offensive content?
    \answerYes{Yes. See \S\ref{sec:ethics}.}
\item If you are curating or releasing new datasets, did you discuss how you intend to make your datasets FAIR (see \citet{fair})?
\answerNA{NA.}
\item If you are curating or releasing new datasets, did you create a Datasheet for the Dataset (see \citet{gebru2021datasheets})? 
\answerNA{NA.}
\end{enumerate}

\item Additionally, if you used crowdsourcing or conducted research with human subjects, \textbf{without compromising anonymity}...
\begin{enumerate}
  \item Did you include the full text of instructions given to participants and screenshots?
    \answerNA{NA.}
  \item Did you describe any potential participant risks, with mentions of Institutional Review Board (IRB) approvals?
    \answerNA{NA.}
  \item Did you include the estimated hourly wage paid to participants and the total amount spent on participant compensation?
    \answerNA{NA.}
   \item Did you discuss how data is stored, shared, and deidentified?
   \answerNA{NA.}
\end{enumerate}

\end{enumerate}

\section*{Ethics Statement}\label{sec:ethics}

This study involves the analysis of online data from extremist and general social media forums, raising ethical considerations related to privacy, consent, and the potential impact of the research. All data used in this study were collected from publicly available sources where users posted content in public forums. No private or personally identifiable information was accessed or used in the analysis. To mitigate any potential harm, we took careful steps to anonymize the data, and user-level analyses were conducted with caution, ensuring that no individual participants could be identified.

The potential risks of using this data for research include the reinforcement of stigmas surrounding certain groups, as well as unintended consequences that may arise from the misuse of findings. To address this, we approached the analysis from a neutral, empirical perspective, aiming to better understand patterns of radicalization while being mindful of the broader social context in which this information is used.

Additionally, we acknowledge that studying extremist discourse may inadvertently provide insights into how these groups operate, which could be used for harmful purposes. To prevent this, we focused on developing models that contribute to early detection and intervention strategies, with the aim of mitigating the harm caused by radicalization. We are committed to ensuring that the findings of this work are used to support efforts in counter-radicalization, public safety, and mental health intervention.

This research was conducted in line with institutional guidelines for the ethical use of publicly available data and followed the principles of responsible AI development, particularly regarding the societal impact of extremism detection and intervention strategies.

\appendix

\section*{Appendix}
\label{sec:appendix}

\section{Term Definitions and Disambiguation}
\label{app:term_def}

\textbf{Extremism} is the promotion or advancement of an ideology based on violence, hatred, or intolerance that aims to negate fundamental rights, undermine democratic systems, or create permissive environments for such actions~\cite{extremism2024govuk}. It represents views and behaviors that are not just \textit{unconventional}, but rather far removed from mainstream societal norms and attitudes, advocating for extreme measures that threaten democratic systems and the ability of people to live equally under the law. The manifestation of extremist beliefs spans a spectrum, ranging from subtle expressions of intolerance to overt displays of hatred, ultimately culminating in explicit acts of violence against marginalized groups~\cite{schmid2022violent, knight2017violent}.

\textbf{Radicalization} is the \textit{process} of change by which an individual or group comes to adopt increasingly extreme views in opposition to a political, social, economic, or religious status quo~\cite{police2009radicalization}. This could result in explicit outcomes such as violence or hate-mongering, leading to the degeneration of the individual or the group into extremist ideologies. Radicalization and extremist tendencies usually go hand-in-hand. However, while radicalization focuses more on developing beliefs advocating for a drastic transformation of the social structure, tending towards more and more extreme measures, extremism comprises not just the ideologies but intentional actions and behaviors to propagate those ideologies as well~\cite{trip2019psychological}. 

\textbf{Hate speech} is public speech or texts that encourage violence against specific social groups based on race, ethnicity, religion, etc.~\cite{hatespeechdef}.
While hate speech is commonly studied in NLP contexts~\cite{macavaney2019hate}, studying extremism and radicalization online goes beyond just looking for hate speech or speech inciting violence. Although hate speech may stem from deep-rooted extremist beliefs~\cite{bilewicz2020hate}, it can also manifest from targeted hostility toward individuals based on personal grievances unrelated to their social identity.~\cite{fortuna-etal-2020-toxic,piot2024metahate}
This distinction highlights that hate speech, while potentially overlapping with extremist ideologies and radicalization processes, does not necessarily indicate their presence. 
Our method is comprehensive -- it goes beyond language-based violence incitement to examine fundamental aspects such as underlying beliefs, documented belief transitions, personality traits, dogmatic tendencies, and political, religious, and economic orientations, all grounded in extensive academic research to accurately identify extremist inclinations and radicalization patterns.

\section{Banned Subreddit Case Study}\label{app:subreddits}

\begin{tabular}{p{0.44\linewidth} p{0.47\linewidth}}
\textbf{Banned}
\begin{itemize}
    \item \textit{r/Right\_Wing\_Politics}
    \item \textit{r/europeannationalism}
    \item \textit{r/Identitarians}
    \item \textit{r/DebateAltRight}
    \item \textit{r/AltRightChristian}
    \item \textit{r/pol}
    \item \textit{r/NationalSocialism}
    \item \textit{r/AgainstTheChimpire}
\end{itemize}
&
\textbf{Not-Banned}
\begin{itemize}
    \item \textit{r/AmericanPolitics}
    \item \textit{r/Classical\_Liberals}
    \item \textit{r/gunpolitics}
    \item \textit{r/CanadianPolitics}
    \item \textit{r/JordanPeterson}
    \item \textit{r/justicedemocrats}
    \item \textit{r/StillSandersForPres}
    \item \textit{r/CapitalismVSocialism}
\end{itemize}
\end{tabular}

\section{Compute and Hyperparameters}\label{sec:compute}

We run our experiments on an NVIDIA-RTX-A6000 with 50 GB of memory in an internal server.

\begin{table*}[]
\small
\begin{tabular}{p{0.18\linewidth}p{0.47\linewidth}p{0.04\linewidth}p{0.26\linewidth}}
\toprule
\textbf{Factor} & \textbf{Sentences} & \multicolumn{1}{l}{\textbf{Scores}} & \textbf{Questionnaire} \\
\midrule
\multirow{20}{*}{Revolutionary Attitude} & It is better for government leaders to make decisions without consulting anyone. & 0.80 & General Extremist \\
 & People in government must enforce their authority even if it means violating the rights of some citizens. & 0.57 & General Extremist \\
 & Under some circumstances, a nondemocratic government can be preferable. & 1.15 & General Extremist \\
 & A concentration of power in one person guarantees order. & 0.70 & General Extremist \\
 & Most people in this country have a lifestyle and culture that is necessary to change totally. & 0.59 & Extremism Scale \\
 & It is necessary to totally change the economic system that is the basis of society. & 0.96 & Extremism Scale \\
 & Those who think like me have to thoroughly change the foundation of our own life (economy, job, consumption, well-being). The rest of the society can do what they want. & 0.84 & Extremism Scale \\
 & It is necessary to do away with the democratic form of government if we want to have a decent society. & 0.91 & Extremism Scale \\
 & Just let the rest of the society choose democracy – I, and those who think like me, work to establish up a different system in our own milieu. & 0.89 & Extremism Scale \\
 & There is only one way to live the good and correct life. & 0.58 & Extremism Scale \\
 & A decent living is only possible with socialism. & 0.99 & Left-Wing Radical \\
 & Capitalism is ruining the world. & 0.53 & Left-Wing Radical \\
 & Fascism shows the true face of capitalism. & 0.59 & Left-Wing Radical \\
 & National states should be abolished. & 0.67 & Left-Wing Radical \\
 & The old-fashioned ways and old-fashioned values still show the best way to live. & 0.72 & Right-Wing Authoritarianism \\
 & The society needs to show openness towards people thinking differently, rather than a strong leader, the world is not particularly evil or dangerous. & -0.60 & Right-Wing Authoritarianism \\
 & An ideal society requires some groups to be on top and others to be on the bottom. & 0.76 & Social Dominance Orientation \\
 & It would be best if every people also had its own state. & 0.90 & Ethnic Intolerance \\
 & One should only help other countries if this is to the advantage of one’s own country. & 0.69 & Nationalism Scale \\
 & If other countries accepted more of what we do here, they would be better off & 0.72 & Nationalism Scale \\
 \midrule
\multirow{14}{*}{Radical and Violent Intent} & The (nationality) foreign policy is racist. & 0.51 & Left-Wing Radical \\
 & I would attack police or security forces if I saw them beating members of my ethnic group. & 0.79 & Radicalism Intention \\
 & I would continue to support a group that fights for my ethnic group’s political and legal rights even if the group sometimes breaks the law. & 0.89 & Radicalism Intention \\
 & I would continue to support a group that fight’s for my ethnic group’s political and legal rights even if the groups sometimes resorts to violence. & 0.94 & Radicalism Intention \\
 & I would participate in a public protest against oppression of my ethnic group even if I thought the protest might turn violent. & 0.71 & Radicalism Intention \\
 & As a last resort I’m personally ready to use violence for the sake of my ethnic group. & 0.93 & Violent Intention \\
 & If nothing else helps, I’m prepared to use violence to defend my ethnic group. & 0.95 & Violent Intention \\
 & I’m ready to go and fight for my ethnic group in another country. & 0.96 & Violent Intention \\
 & I will not personally use violence to help my ethnic group. & -0.86 & Violent Intention \\
 & I will personally use violence against people harming other ethnic group members that I care about. & 0.86 & Violent Intention \\
 & Even as a last resort, I will not use violence for the sake of other ethnic group members. & -0.92 & Violent Intention \\
 & My ethnic group is an important reflection of who I am. & 0.53 & Self-Categorization Scale \\
 & I identify with other members of my ethnic group. & 0.53 & Self-Categorization Scale \\
 & I am like other members of my ethnic group & 0.52 & Self-Categorization Scale \\
 \midrule 
 \end{tabular}
\end{table*}

\begin{table*}
\small 
\begin{tabular}{p{0.25\linewidth}p{0.40\linewidth}p{0.04\linewidth}p{0.25\linewidth}}
\toprule
\textbf{Factor} & \textbf{Sentences} & \multicolumn{1}{l}{\textbf{Scores}} & \textbf{Questionnaire} \\
\midrule
\multirow{12}{*}{Social Dominance Orientation} & It is a waste of time to try to find common solutions with those whose thoughts about life are completely different than ours. & 0.63 & Extremism Scale \\
 & Jews simply have something special and peculiar about them and do not really fit in with us. & 0.70 & Right-Wing Radical \\
 & Actually, (country) are inherently superior to other people. & 0.55 & Right-Wing Radical \\
 & We should work to give all groups an equal chance to succeed. & -0.76 & Social Dominance Orientation \\
 & An ideal society requires some groups to be on top and others to be on the bottom. & 0.50 & Social Dominance Orientation \\
 & Some groups of people are simply inferior to other groups. & 1.06 & Social Dominance Orientation \\
 & No one group should dominate in society. & -0.61 & Social Dominance Orientation \\
 & Groups at the bottom are just as deserving as groups at the top. & -0.90 & Social Dominance Orientation \\
 & Group equality should not be our primary goal. & 0.86 & Social Dominance Orientation \\
 & It is unjust to try to make groups equal. & 0.96 & Social Dominance Orientation \\
 & We should do what we can to equalize conditions for different groups. & -0.62 & Social Dominance Orientation \\
 & It is better if only members of the same people get married to each other. & 0.63 & Ethnic Intolerance \\
 \midrule
\multirow{12}{*}{Cold and Calculating} & I tend to manipulate others to get my way. & 0.85 & Dirty Dozen \\
 & I have used deceit or lied to get my way. & 0.90 & Dirty Dozen \\
 & I have used flattery to get my way. & 0.93 & Dirty Dozen \\
 & I tend to exploit others towards my own end. & 0.89 & Dirty Dozen \\
 & I tend to lack remorse. & 0.66 & Dirty Dozen \\
 & I tend to be unconcerned with the morality of my actions. & 0.61 & Dirty Dozen \\
 & I tend to be callous or insensitive. & 0.73 & Dirty Dozen \\
 & I tend to be cynical. & 0.74 & Dirty Dozen \\
 & I tend to want others to admire me. & 0.78 & Dirty Dozen \\
 & I tend to want others to pay attention to me. & 0.80 & Dirty Dozen \\
 & I tend to seek prestige or status. & 0.76 & Dirty Dozen \\
 & I tend to expect special favors from others & 0.86 & Dirty Dozen \\
 \midrule
\multirow{9}{*}{State control} & The government should close communication media that are critical. & 0.83 & General Extremist \\
 & The persecution of and spying on left-wing system critics by the state and police is increasing. & 0.66 & Left-Wing Radical \\
 & Our country needs a powerful leader, in order to destroy the radical and immoral currents prevailing in society today. & 0.50 & Right-Wing Authoritarianism \\
 & God's laws about abortion, pornography and marriage must be strictly followed before it is too late, violations must be punished. & 0.79 & Right-Wing Authoritarianism \\
 & It would be best if newspapers were censored so that people would not be able to get hold of destructive and disgusting material. & 0.92 & Right-Wing Authoritarianism \\
 & People ought to put less attention to the Bible and religion, instead they ought to develop their own moral standards. & -0.52 & Right-Wing Authoritarianism \\
 & There are many radical, immoral people trying to ruin things; the society ought to stop them. & 0.64 & Right-Wing Authoritarianism \\
 & It is better to accept bad literature than to censor it. & -0.73 & Right-Wing Authoritarianism \\
 & Facts show that we have to be harder against crime and sexual immorality, in order to uphold law and order. & 0.72 & Right-Wing Authoritarianism \\
 \midrule
\end{tabular}%

\end{table*}

\begin{table*}
\small
\begin{tabular}
{p{0.25\linewidth}p{0.40\linewidth}p{0.04\linewidth}p{0.25\linewidth}}
\toprule
\textbf{Factor} & \textbf{Sentences} & \multicolumn{1}{l}{\textbf{Scores}} & \textbf{Questionnaire} \\
\midrule
\multirow{5}{*}{Moral Detachment} & If one does not live in agreement with the good and correct life, then one has chosen to withdraw from the community. & 0.54 & Extremism Scale \\
 & It is a waste of time to try to find common solutions with those whose thoughts about life are completely different than ours. & 0.58 & Extremism Scale \\
 & I’m not prepared to use violence in any situation. & -0.66 & Violent Intention \\
 & I tend to lack remorse. & 0.52 & Dirty Dozen \\
 & I tend to be unconcerned with the morality of my actions. & 0.60 & Dirty Dozen \\
 \midrule
\multirow{3}{*}{Nationalism} & We should have the courage to have a strong sense of national consciousness. & 0.61 & Right-Wing Radical \\
 & It is the foremost duty of each young American to honor the national history and its heritage. & 0.63 & Nationalism Scale \\
 & Because of our important historical experience, we should have more to say in international affairs. & 0.79 & Nationalism Scale \\
 \midrule
\multirow{4}{*}{Dogmatism} & There is only one way to live the good and correct life. & 0.49 & Extremism Scale \\
 & God's laws about abortion, pornography and marriage must be strictly followed before it is too late, violations must be punished. & 0.79 & Right-Wing Authoritarianism \\
 & People ought to put less attention to the Bible and religion, instead they ought to develop their own moral standards. & -0.37 & Right-Wing Authoritarianism \\
 & Facts show that we have to be harder against crime and sexual immorality, in order to uphold law and order. & 0.46 & Right-Wing Authoritarianism \\
 \midrule
\multirow{3}{*}{Anti-capitalism} & A decent living is only possible with socialism. & 0.46 & Left-Wing Radical \\
 & Capitalism is ruining the world. & 0.66 & Left-Wing Radical \\
 & Fascism shows the true face of capitalism. & 0.72 & Left-Wing Radical \\
 \midrule
\multirow{3}{*}{Xenophobia} & The (country) has become too foreign to a dangerous extent due to all the foreigners here. & 0.80 & Right-Wing Radical \\
 & Foreigners and asylum seekers are the ruin of (country). & 0.59 & Right-Wing Radical \\
 & If there are too many foreigners in the country, one might as well let them feel that they are not welcome. & 0.52 & Ethnic Intolerance \\
 \midrule
\multirow{3}{*}{National Self-interest} & The (nationality) foreign policy is racist. & 0.30 & Left-Wing Radical \\
 & One should only help other countries if this is to the advantage of one’s own country. & 0.68 & Nationalism Scale \\
 & If other countries accepted more of what we do here, they would be better off. & 0.39 & Nationalism Scale\\
 \bottomrule
\end{tabular}
\caption{Factors of the ``Extremist Eleven'' along with each factor's highest scoring extremism scale items.}
\label{tab:extremist_eleven}
\end{table*}

\end{document}